\documentclass{article}

\usepackage{amsmath}

\newcommand{\be}{\begin{equation}}
\newcommand{\ee}{\end{equation}}
\newcommand{\ben}{\begin{eqnarray}}
\newcommand{\een}{\end{eqnarray}}

\newcommand{\nd}{\noindent}

\title{{\bf Possible Divergences in Tsallis' Thermostatistics}}
\author{A. Plastino, M. C. Rocca\\
La Plata National University\\
and Argentina's National Research Council,\\
(IFLP-CCT-CONICET)-C. C. 727, 1900\\
La Plata - Argentina}

\date{\today}

\begin{document}

\maketitle

\begin{abstract}

\nd Trying to compute the nonextensive q-partition function for
the Harmonic Oscillator in more than two dimensions, one
encounters that it diverges, which poses a serious threat to
Tsallis' thermostatistics. Appeal to the so called q-Laplace
Transform, where the q-exponential function plays the role of the
ordinary exponential, is seen to save the day.

PACS {05.20.-y, 05.70.Ce, 05.90.+m}

\end{abstract}

\section{Introduction}

Divergences are quite important in theoretical physics. Indeed,
the study and elimination of divergences of a physical theory is
perhaps one of the most important branches of theoretical physics.
The quintessential typical example is the attempt to quantify the
gravitational field, which so far has not been achieved. Some
examples of elimination of divergences can be seen in  references
\cite{tq1,tq2,tq3,tq4,tq5}.

\nd The so-called q-exponential function \cite{tsallis}

\ben & e_q(x)=  [1+ (1-q) x]_+^{1/(1-q)}; \cr &   q \in
\mathcal{R}; \,\, e_q(x)\rightarrow e^{x} \,\,{\rm when}\,\, q
\rightarrow 1, \een is the flagship of non-extensive statistics
(see \cite{tsallis} and references therein), a subject that has
captured the interest of literally hundreds of researchers, that
have produced several thousand papers in such respect in the last
years \cite{web}. Indeed,  natural phenomena and  laboratory
experiments yield  a wide spectrum of empiric results
demonstrating data-distributions clearly deviating from
exponential decay \cite{tsallis,web}. Non-extensive statistical
mechanics  is an approach that  explains this non-Boltzmann
behavior, with deformed exponential distributions (such as
q-exponentials exhibiting long tails when $q > 1$). These
distributions are  empirically encountered in a variety of
scientific disciplines. One can mention  subjects as variegated as
 turbulence,  cosmic rays, earthquake's
magnitudes' distributions, speed distributions in bacterial
populations, geological, nuclear, particle,  and cosmic phenomena,
or financial market data \cite{tsallis,web}. \vskip 3mm

\nd Moreover, the $e_q-$functions are the natural solutions to an
interesting  new version of the nonlinear Schr\"odinger equation
(NLSE), recently advanced by Nobre, Rego-Monteiro and Tsallis
\cite{NRT11,NRT12} (see also \cite{Curilef}). This NLSE
constitutes  an intriguing proposal that is part of a program to
investigate non-linear versions of some of the basic equations of
physics, a research venue that registers significant activity
\cite{S07,SS99}. Here we show that, when regarded as a probability
distribution function, the q-exponential leads to a divergent
partition function in two or more dimensions, which constitutes a
potential catastrophe for q-nonextensivity, with several thousands
of papers referring to it in the last 15 years.

\nd One should mention that Boon and Lutsko  \cite{boon, boon1}
have already shown, in two interesting papers, that divergences
exist in Tsallis' thermo-statistics in some classical settings.
\vskip 3mm \nd \fbox{\parbox{4.6in}{What we discuss here is {\it
how to avoid these divergences} in Tsallis' theory both for the
harmonic oscillator and in the general case of a well behaved
Hamiltonian. Our main idea revolves around the concept of energy
density, central in statistical mechanics, as seen for example in
the classical text-book by Reif \cite{reif}. }}
\section{Partition Function for the Harmonic Oscillator (HO)}
\nd Using appropriate units, the partition function of
n-dimensional harmonic oscillator is
\begin{equation}
\label{ep1}
{\cal Z}=\int\limits_{-\infty}^{\infty}
e^{-\beta (P^2+Q^2)} \;d^np\;d^nq
\end{equation}
where
$P^2=p_1^2+p_2^2+\cdot\cdot\cdot p_n^2$,
$Q^2=q_1^2+q_2^2+\cdot\cdot\cdot q_n^2$\\
Taking into account that i) the value of a solid angle in
n-dimensional $p$-space and $q$-space is (see \cite{tp1})
$\varOmega_p=\varOmega_q=2\pi^{n/2}/\Gamma(n/2)$, ii) performing
the change of variables $P^2+Q^2=U$ $Q=\sqrt{U-P^2}$, and iii)
using the result
\begin{equation}
\label{ep2} \int\limits_0^u x^{\nu-1} (u-x)^{\mu-1}\;dx=
u^{\mu+\nu-1}{\cal B}(\mu,\nu),
\end{equation}
where ${\cal B}$ is the Euler's Beta function \cite{tp2}, we
obtain for ${\cal Z}$
\begin{equation}
\label{ep3} {\cal Z}=\frac {\pi^n} {\Gamma(n)}
\int\limits_0^{\infty} U^{n-1}e^{-\beta U} \;dU,
\end{equation}
 $U$ being the HO-energy  and
$g(U)=[\pi^n/\Gamma(n)] U^{n-1}$  the associated energy density.

\vskip 3mm \nd \fbox{\parbox{4.6in}{It is of the essence to note
that the partition function can also be obtained as the Laplace
Transform of the energy density \cite{jchem}.}}\vskip 3mm

\nd  Following a similar line of reasoning to that leading to
(\ref{ep3}), we obtain for the {\it mean energy} ${\cal U}$ and
the entropy $S$
\begin{equation}
\label{ep4} {\cal U}=\int\limits_{-\infty}^{\infty} (P^2+Q^2)
\frac {e^{-\beta (P^2+Q^2)}} {{\cal
Z}}\;d^np\;d^nq\,\,\,\,\Rightarrow
\end{equation}
\begin{equation}
\label{ep5} {\cal U}=\frac {\pi^n} {\Gamma(n){\cal Z}}
\int\limits_0^{\infty} U^n e^{-\beta U} \;dU, \,\,\,\,{\rm and}
\end{equation}

\begin{equation}
\label{ep6} {\cal S}=\frac {\pi^n} {\Gamma(n){\cal Z}}
\int\limits_0^{\infty} (\ln{\cal Z}+\beta U)U^{n-1} e^{-\beta U}
\;dU.
\end{equation}

\section{Divergences in  Tsallis' Theory}
\nd Lutsko and Boon have discussed divergences in Tsallis' theory
\cite{boon, boon1}. We demonstrate the same below in what we
believe is a more direct, straightforward, and explicit fashion.
In the nonextensive (Tsallis') approach the corresponding values
for the q-partition function ${\cal Z}_q$, the mean energy ${\cal
U}$, and the Tsallis' entropy ${\cal S}$ are obtained using
\cite{tsallis}: i) the q-exponential function of energy instead of
the exponential function and ii) the q-logarithm function in place
of the logarithmic  function. One has, for the probability,
\begin{equation}
\label{et1} P_q[H(p,x)]=\frac {e_q[-\beta H(p,x)]} {{\cal Z}_q},
\end{equation}
where
\begin{equation}
\label{et2} {\cal Z}_q=\int e_q[-\beta H(p,x)]\;d^npd^nx,
\end{equation}
while the mean energy is
\begin{equation}
\label{et3} {\cal U}=\int H(p,x)P_q[H(p,x)]\;d^npd^nx,
\end{equation}
and  the entropy:
\begin{equation}
\label{et4} S=-\int P_q[H(p,x)] \ln_q P_q[H(p,x)]\; d^npd^nx,
\end{equation}
with
\begin{equation}
\label{et5}
\ln_q x=\frac {x^{1-q}-1} {1-q}\;\rightarrow \ln x\;\;\;for\;\;\; q\rightarrow 1
\end{equation}
One then finds
\begin{equation}
\label{ep7} {\cal Z}_q=\frac {\pi^n} {\Gamma(n)}
\int\limits_0^{\infty}U^{n-1} [1+(q-1)\beta U]^{\frac {1} {1-q}}
\;dU,
\end{equation}
where the real parameter $q$ obeys $1<q<2$, and
\begin{equation}
\label{ep8} {\cal U}=\frac {\pi^n} {\Gamma(n){\cal Z}_q}
\int\limits_0^{\infty}U^n [1+(q-1)\beta U]^{\frac {1} {1-q}} \;dU,
\end{equation}
\[{\cal S}=\left\{\frac {\pi^n({\cal Z}_q^{1-q}-1)} {\Gamma(n){\cal Z}_q^{2-q}(1-q)}
\int\limits_0^{\infty}U^{n-1}
[1+(q-1)\beta U]^{\frac {1} {1-q}} \;dU +\right.\]
\begin{equation}
\label{ep9} \left.\frac {\pi^n\beta} {\Gamma(n){\cal Z}_q^{2-q}}
\int\limits_0^{\infty}U^n [1+(q-1)\beta U]^{\frac {1} {1-q}}
\;dU\right\}.
\end{equation}
Looking at (\ref{ep7}), we immediately detect a serious problem:
the partition-defining integral  diverges for   $q\geq 3/2$ and
$n\geq 2$. For example, if $q=3/2$ and $n\geq 2$ we have
\begin{equation}
\label{ep10} {\cal Z}_q=\frac {\pi^n} {\Gamma(n)}
\int\limits_0^{\infty}U^{n-1} \left[1+\frac {\beta U}
{2}\right]^{-2} \;dU,
\end{equation}
which is clearly divergent. For the average energy the situation is even
worse.

\vskip 3mm \nd \fbox{\parbox{4.6in}{For $q\geq 3/2$ and $n\geq 1$
we see that the integral is divergent, even in the one-dimensional
case.}}

\vskip 2mm \nd For example for $q=3/2$ and $n\geq 1$ we obtain:
\begin{equation}
\label{er1} {\cal U}=\frac {\pi^n} {\Gamma(n)}
\int\limits_0^{\infty}U^n \left[1+\frac {\beta U}
{2}\right]^{-2} \;dU,
\end{equation}
This integral is divergent. The integral (\ref{ep9}) registers a
similar pitfall.

\section{Solution via  q-Laplace Transforms of the energy density \cite{tp3}}
\nd The origin of these divergences that, as  we have just
demonstrated, plague Tsallis' theory, is clear. It is well known
(Cf.  (\ref{ep3})) that ${\cal Z}$ is the Laplace Transform of the
energy density. Thus, (\ref{ep7}) should be the q-Laplace
Transform \cite{tp3} (that replaces the q-exponential function) of
the energy density, but this is not so. Accordingly, the correct
way of obtaining a ${\cal Z}_q$ should pass through the q-Laplace
Transform of the energy density, as explained at length in
\cite{tp3}. \vskip 3mm

\nd In reference \cite{tp3}  one sees that the expression for the
unilateral q-Laplace transform of a function $f(U)\in\Omega_I$
reads

\begin{equation}
\label{eq1} L(\beta,q)=H[\Re(\beta)]\int\limits_0^{\infty}
f(U)\{1-(1-q)\beta U [f(U)]^{(q-1)}\}^{\frac {1} {1-q}}\;dU,
\end{equation}
where $H$ is the Heaviside function and {\it the brackets
correspond to the argument of the q-Laplace transform, that will
play a leading role below, for the special function
$f(U)=U^{n-1}$} (for a definition of $\Omega_I$ see \cite{tp3}).
Consequently, ${\cal Z}_q$ should be evaluated via
\begin{equation}
\label{ep11} {\cal Z}_q=\frac {\pi^n} {\Gamma(n)}
\int\limits_0^{\infty}U^{n-1} [1+(q-1)\beta
U^{(n-1)(q-1)+1}]^{\frac {1} {1-q}} \;dU,
\end{equation}
and, in similar fashion, for ${\cal U}$ and ${\cal S}$ one should
have the  expressions
\begin{equation}
\label{ep12} {\cal U}=\frac {\pi^n} {\Gamma(n){\cal Z}_q}
\int\limits_0^{\infty}U^n [1+(q-1)\beta U^{n(q-1)+1}]^{\frac {1}
{1-q}} \;dU,
\end{equation}
\[{\cal S}=\left\{\frac {\pi^n({\cal Z}_q^{1-q}-1)}
{\Gamma(n){\cal Z}_q^{2-q}(1-q)}\times\right.\]
\[\int\limits_0^{\infty}U^{n-1}
[1+(q-1)\beta U^{(n-1)(q-1)+1}]^{\frac {1} {1-q}} \;dU\]
\begin{equation}
\label{ep13} \left.+\frac {\pi^n\beta} {\Gamma(n){\cal Z}_q^{2-q}}
\int\limits_0^{\infty}U^n [1+(q-1)\beta U^{n(q-1)+1}]^{\frac {1}
{1-q}} \;dU\right\}.
\end{equation}
Integral (\ref{ep11}) can be evaluated making the change of
variable $x=U^{(n-1)(q-1)+1}$ and using (see Ref. \cite{tp4})
\begin{equation}
\label{ep14} \int\limits_0^{\infty} \frac {x^{\mu-1}} {(1+\beta
x)^{\nu}}\;dx= \beta^{-\mu}{\cal B}(\mu,\nu-\mu),
\end{equation}
that leads us to find for  ${\cal Z}_q$ a convergent expression,
namely,
\[{\cal Z}_q=\left\{\frac {\pi^n[\beta(q-1)]^{-\frac {n} {(n-1)(q-1)+1}}}
{\Gamma(n)[(n-1)(q-1)+1]}\times\right.\]
\begin{equation}
\label{ep15} \left.{\cal B}\left[\frac {n} {(n-1)(q-1)+1},\frac
{1} {q-1}- \frac {n} {(n-1)(q-1)+1}\right]\right\}.
\end{equation}
Analogously, we find convergent expressions for $U$ and $S$
\[{\cal U}=\left\{\frac {\pi^n[\beta(q-1)]^{-\frac {n+1} {n(q-1)+1}}}
{\Gamma(n){\cal Z}_q[n(q-1)+1]}\times\right.\]
\begin{equation}
\label{ep16}
\left.{\cal B}\left[\frac {n+1} {n(q-1)+1},\frac {1} {q-1}-
\frac {n+1} {n(q-1)+1}\right]\right\}
\end{equation}
and
\[{\cal S}=\left\{\frac {\pi^n({\cal Z}_q^{1-q}-1)[\beta(q-1)]^{-\frac {n} {(n-1)(q-1)+1}}}
{\Gamma(n){\cal Z}_q^{2-q}(1-q)[(n-1)(q-1)+1]}\times\right.\]
\[\left.{\cal B}\left[\frac {n} {(n-1)(q-1)+1},\frac {1} {q-1}-
\frac {n} {(n-1)(q-1)+1}\right]\right\}+\]
\[\left\{\frac {\pi^n\beta[\beta(q-1)]^{-\frac {n+1} {n(q-1)+1}}}
{\Gamma(n){\cal Z}_q^{2-q}[n(q-1)+1]}\times\right.\]
\begin{equation}
\label{ep17}
\left.{\cal B}\left[\frac {n+1} {n(q-1)+1},\frac {1} {q-1}-
\frac {n+1} {n(q-1)+1}\right]\right\}
\end{equation}
We appreciate thus that the use of the q-Laplace Transform of the
energy density  makes all $q-$thermodynamics' variables to be
finite.

\section{The General Case}

\nd  In the general case of a Hamiltonian which depends on $2n$
variables $p_1,p_2,...,p_n$ and $ q_1,q_2,...,q_n$ we have
\begin{equation}
\label{er2}
{\cal Z}=\int\limits_{-\infty}^{\infty}
e^{-\beta H(p,q)} \;d^np\;d^nq
\end{equation}
Appealing, for example, to the change of variables
$U=H(p,q)$,\vskip 2mm \nd  $q_i=g(U,
p_1,...,p_n,q_1,...,q_{i-1},q_{i+1},...,q_n)$ we obtain for ${\cal
Z}$
\begin{equation}
\label{er3} {\cal Z}=\int\limits_0^{\infty}e^{-\beta U}\;dU
\int\limits_{-\infty}^{\infty} J(U,
p,q_1,...,q_{i-1},q_{i+1},...,q_n) \;d^np\;d^{n-1}q.
\end{equation}
where $J$ is the Jacobian of the change of variables, that yields
an ``energy density". We then obtain for this energy density $f$
the expression
\begin{equation}
\label{er4} f(U)= \int\limits_{-\infty}^{\infty} J(U,
p,q_1,...,q_{i-1},q_{i+1},...,q_n) \;d^np\;d^{n-1}q.
\end{equation}
Thus,  ${\cal Z}$ can be written in the form:
\begin{equation}
\label{er5} {\cal Z}=\int\limits_0^{\infty}f(U) e^{-\beta U} \;dU.
\end{equation}
Analogously we obtain for ${\cal U}$ and $S$
\begin{equation}
\label{er6} {\cal U}=\frac {1} {{\cal
Z}}\int\limits_0^{\infty}Uf(U) e^{-\beta U} \;dU.
\end{equation}
\begin{equation}
\label{er7} S=\frac {1} {{\cal Z}}\int\limits_0^{\infty} (\ln
{\cal Z}+\beta U)f(U) e^{-\beta U} \;dU.
\end{equation}
Considering that for a well behaved Hamiltonian $f(U)$ {\it is an
analytic function in the upper right quadrant of the complex
plane} we are entitled to  write
\begin{equation}
\label{er8} f(U)=\sum\limits_{n=0}^{\infty}a_n U^n,
\end{equation}
and obtain the convergent result
\begin{equation}
\label{er9} {\cal Z}=\sum\limits_{n=0}^{\infty}
a_n\int\limits_0^{\infty} U^n e^{-\beta U} \;dU,
\end{equation}
\begin{equation}
\label{er10} {\cal U}=\frac {1} {{\cal Z}}
\sum\limits_{n=0}^{\infty} a_n\int\limits_0^{\infty} U^{n+1}
e^{-\beta U} \;dU,
\end{equation}
\begin{equation}
 \label{er11}
S=\frac {1} {{\cal Z}} \ln {\cal Z} \sum\limits_{n=0}^{\infty}
a_n\int\limits_0^{\infty} U^n e^{-\beta U} \;dU + \frac {\beta}
{{\cal Z}} \sum\limits_{n=0}^{\infty} a_n\int\limits_0^{\infty}
U^{n+1} e^{-\beta U} \;dU.
\end{equation}
Taking into account the nonlinearity of the q-Laplace transform,
from (\ref{er9}),(\ref{er10}), and (\ref{er11}) we obtain, {\it
adapting things to the nonextensive, q-scenario},
\begin{equation}
\label{er12} {\cal Z}_q= \sum\limits_{n=0}^{\infty}a_n
\int\limits_0^{\infty}U^n [1+(q-1)\beta U^{n(q-1)+1}]^{\frac {1}
{1-q}} \;dU,
\end{equation}
\begin{equation}
\label{er13} {\cal U}_q=\frac {1} {{\cal Z}_q}
\sum\limits_{n=0}^{\infty}a_n \int\limits_0^{\infty}U^{n+1}
[1+(q-1)\beta U^{(n+1)(q-1)+1}]^{\frac {1} {1-q}} \;dU,
\end{equation}
\[{\cal S}=\left\{\frac {{\cal Z}_q^{1-q}-1} {{\cal Z}_q^{2-q}(1-q)}
\sum\limits_{n=0}^{\infty} a_n
\int\limits_0^{\infty}U^n
[1+(q-1)\beta U^{n(q-1)+1}]^{\frac {1} {1-q}} \;dU \right.\]
\begin{equation}
\label{er14} \left.+\frac {\beta} {{\cal Z}_q^{2-q}}
\sum\limits_{n=0}^{\infty} a_n \int\limits_0^{\infty}U^{n+1}
[1+(q-1)\beta U^{(n+1)(q-1)+1}]^{\frac {1} {1-q}} \;dU\right\},
\end{equation}
and attain convergence in every instance. Note that having i) a
partition function, ii) a mean energy and iii) an entropy, we
automatically get a thermo-statistics. This may not be the
orthodox one, but is quite a legitimate one nonetheless.

\vskip 3mm

\nd Note that, so as to obtain the entropy (\ref{er14}) we have
started our considerations from the Tsallis entropy definitions
and then we proceeded with our
q-Laplace Transform.

\vskip 3mm \nd \fbox{\parbox{4.6in}{The essence of our maneuvers
was to replace the $q$-exponential by the argument of the
$q$-Laplace transform.  Thus, the center of gravity is displaced
from probability distributions to energy densities. Note that the
later are well-established empirical quantities characterizing a
given system, while the interpretation of the former is a matter
of controversy, as, for instance, Bayesian vs. frequentist. Thus,
this shifting is empirically sound.}}

\section{Conclusions}

\nd It is well known that, for obtaining the partition function
$\cal{Z}$, two alternative routes can be followed:

\begin{itemize}

\item the ``natural" one, given by $\cal{Z}$'s definition and

\item $\cal{Z}$   as the Laplace Transform of the energy density.

\end{itemize}

\nd In the orthodox Boltzmann-Gibbs instance, that uses the
ordinary exponential function, the two routes yield the same
result.\vskip 3mm

\nd We have here proved that such is NOT the case for Tsallis'
thermostatistics, for which the first alternative diverges in
three or more dimensions, due to the long tail of the
q-exponential function.  One must necessarily follow the second
path, that yields finite results. Thus, the q-Laplace Transform
becomes an indispensable tool for nonextensive statistics.

\section*{Acknowledgements}

\nd The authors acknowledge support from CONICET (Argentine
Agency).


\begin{thebibliography}{}

\bibitem{tq1} C. G. Bollini and J. J. Giambiagi: Phys. Lett.
{\bf B 40}, (1972), 566.Il Nuovo Cim. {\bf B 12}, (1972), 20.
\bibitem{tq2} C. G. Bollini and J.J Giambiagi :
Phys. Rev. {\bf D 53}, (1996), 5761.
\bibitem{tq3} G.'t Hooft and M. Veltman: Nucl. Phys. {\bf B 44},
(1972), 189.
\bibitem{tq4} C. G. Bollini, T. Escobar and M. C. Rocca :
Int. J. of Theor. Phys. {\bf 38}, (1999), 2315.
\bibitem{tq5} C. G Bollini and M.C. Rocca :
Int. J. of Theor. Phys. {\bf 43}, (2004), 59.
Int. J. of Theor. Phys. {\bf 43}, (2004), 1019.
Int. J. of Theor. Phys. {\bf 46}, (2007), 3030.

\bibitem{tsallis}  M. Gell-Mann and C. Tsallis, Eds. {\it Nonextensive Entropy:
Interdisciplinary applications}, Oxford University Press, Oxford,
2004;  C. Tsallis, {\it Introduction to Nonextensive Statistical
Mechanics: Approaching a Complex World}, Springer, New York, 2009.

\bibitem{web} See http://tsallis.cat.cbpf.br/biblio.htm for a
regularly updated bibliography on the subject.

\bibitem{NRT11} F.D. Nobre, M A. Rego-Monteiro and C. Tsallis,
Phys. Rev. Lett \textbf{106} (2011) 140601.

\bibitem{NRT12} F.D. Nobre, M.A. Rego-Monteiro, and C. Tsallis,
Europhysics Letters 97 (2012) 41001.

\bibitem{Curilef} S. Curilef, A.R. Plastino, A. Plastino, Physica A {\bf 392} (2013) 2631;
I. V. Toranzo, A. R.  Plastino, J. S. Dehesa, A. Plastino, Physica
A {\bf 392} (2013) 3945.


\bibitem{S07} A.C. Scott, {\it The Nonlinear Universe} (Springer, Berlin, 2007).

\bibitem{SS99}
C. Sulem and P.L. Sulem, {\it The Nonlinear Schrodinger
 Equation: Self-Focusing and Wave Collapse} (Springer,
 New York, 1999).

\bibitem{boon} J. P. Boon and J. F. Lutsko:
Phys. Lett. A {\bf 375}, (2011), 329.

\bibitem{boon1} J. F. Lutsko and J. P. Boon:
Europhys. Lett. {\bf 95}, (2011), 20006.


\bibitem{reif} F. Reif, {\it Fundamentals of statistical and thermal
physics} (McGraw-Hill, NY, 1965).

\bibitem{tp1} I. M. Guelfand and G. E. Chilov, {\it Les
Distributions}, Vol. 1 (Dunod, Paris, 1962).




\bibitem{tp2} I. S. Gradshteyn and I. M. Rizhik, {\it Table
of Integrals Series and Products} (Academic Press, NY, 1965,
p.284, {\bf 3.191},1).


\bibitem{jchem} D. Romanini and K. K. Lehmann, J. Chem. Phys. {\bf 98} (1993)
6437.


\bibitem{tp3} A. Plastino, M.C.Rocca, Physica A {\bf 392}
 (2013) 5581.



\bibitem{tp4} I. S. Gradshteyn and I. M. Rizhik, {\it Table
of Integrals Series and Products} (Academic Press, NY, 1965,
p.284, {\bf 3.194},3).


\end{thebibliography}
\end{document}